\documentclass[aps,prd,twocolumn,showpacs]{revtex4}
\usepackage{graphicx}
\usepackage{dcolumn}
\usepackage{bm}
\usepackage{amsmath}
\usepackage{epstopdf}
\usepackage{amsfonts}
\usepackage{amssymb}%
\providecommand{\U}[1]{\protect\rule{.1in}{.1in}}
\newcommand{\be}{\begin{equation}}
\newcommand{\en}{\end{equation}}
\newcommand{\bea}{\begin{eqnarray}}
\newcommand{\ena}{\end{eqnarray}}
\begin{document}
\title{Observational constraints on warm quasi-exponential inflation}
\author{Nelson Videla}
\email{nelson.videla@pucv.cl}
\affiliation{Instituto de F\'isica, Pontificia Universidad Cat\'olica de Valpara\'iso.\\ Avda. Universidad 330, Curauma, Valpara\'iso, Chile.}
\author{Grigoris Panotopoulos}
\email{grigorios.panotopoulos@tecnico.ulisboa.pt}
\affiliation{CENTRA, Instituto Superior T\'{e}cnico, Universidade de Lisboa, Av. Rovisco Pais 1, 1049-001 Lisboa, Portugal}

\date{\today}

\begin{abstract}
In the present work we study a warm inflationary model defined by a quasi-exponential inflaton potential and an inflaton decay rate proportional to the Hubble rate. The model is characterized by three free parameters. We compute the power spectrum, the scalar spectral index as well as the tensor-to-scalar ratio within the framework of the model, and we compare with the latest Planck data. On the $r-n_s$ plane we show both the theoretical curves and the contour plots allowed by observations, and we constrain the parameters of the model accordingly. The non-linear parameter $f_{NL}$, corresponding to primordial non-Gaussianities, is also discussed and we found that the value predicted by our model is within the bounds imposed by current observational data.
\end{abstract}


\maketitle



\section{Introduction}

Inflation \cite{Starobinsky:1980te,R1,R106,R103,R104,R105,Linde:1983gd} is widely accepted as the standard paradigm of the early Universe. The first reason is due to the fact that several long-standing puzzles of the Hot Big-Bang model, such as the horizon, flatness, and monopole problems, find a natural explanation in the framework of inflationary Universe. In addition, and perhaps the most intriguing feature of inflation, is that it gives us a causal interpretation of the origin of the Cosmic Microwave Background (CMB) temperature anisotropies \cite{astro,astro2,astro202,Hinshaw:2012aka,Ade:2013zuv,Ade:2013uln,Ade:2015xua,Ade:2015lrj}, while at the same time it provides us with a mechanism to explain the Large-Scale Structure (LSS) of the Universe, since quantum fluctuations during the inflationary era may give rise to the primordial density perturbations \cite{Starobinsky:1979ty,R2,R202,R203,R204,R205}.

Standard cold inflation requires two separate phases as follows: First, in the  slow-roll approximation \cite{Lyth:2009zz} the Universe undergoes a dramatic accelerating expansion during which the energy density of the Universe is dominated by a scalar field called the inflaton. Subsequently, during the reheating phase \cite{Kofman:1994rk, Kofman:1997yn,Allahverdi:2010xz,Amin:2014eta} the inflaton oscillates around the minimum of its potential, and the Universe enters the radiation era of the standard Hot Big-Bang model.

The inflationary paradigm is tested and constrained upon comparison to current astrophysical and cosmological observations, in particular those that come from the CMB temperature anisotropies. In practice, the predictions of representative inflationary potentials are given on the $n_s-r$ plane, where the allowed contour plots from the data are also shown. Recently, the Planck collaboration published new more precise data regarding the CMB temperature anisotropies \cite{Ade:2015lrj}. The latest Planck data have improved the upper bound on the tensor-to-scalar ratio $r_{0.002} < 0.11$($95\%$ CL), which is similar to $r < 0.12$ ($95\%$ CL) obtained in \cite{Ade:2013uln}.

Warm inflation is an alternative to standard cold inflation. Contrary to what happens in cold inflation, during which the temperature of the Universe drops dramatically and then a reheating phase is required so that the Universe can enter the radiation era, which is essential for a successful primordial Big-Bang Nucleosynthesis, warm inflation is characterized by the essential feature that after the slow-roll phase, the Universe smoothly enters the radiation era and thus a reheating phase is no longer required \cite{warm1,warm2}. As a matter of fact, several inflationary models excluded by current data in the standard cold inflation scenario, can be rescued in warm inflation thanks to the different dynamics of the new scenario. For a representative list
of recent references see e.g. \cite{Bastero-Gil:2015nja,Panotopoulos:2015qwa,Bastero-Gil:2016qru,Visinelli:2016rhn,Gim:2016uvv,Oyvind Gron:2016zhz,Benetti:2016jhf,Jawad:2017rkq}.

Dissipative effects arise from a friction term (or else dissipative coefficient) $\Gamma$, which describes the processes of the scalar field dissipating into a thermal bath via its interactions with other degrees of freedom. The effectiveness of warm inflation may be parameterized by the ratio $Q \equiv \Gamma/3H$. The weak dissipative regime for warm inflation corresponds to the condition $Q \ll 1$, while $Q \gg 1$ characterizes the strong dissipative regime of warm inflation. It is important to emphasize that the dissipative coefficient $\Gamma$ may be computed from first principles in quantum field theory considering that $\Gamma$ encodes the microscopic physics resulting from the interactions between the inflaton and other fields that may be present \cite{BasteroGil:2012cm,Bartrum:2013fia,Zhang:2009ge,26,28,PRD}.  In general, the inflaton decay rate may depend on the scalar field itself or the temperature of the thermal bath, or both quantities, or even it can be a constant.

What is more, thermal fluctuations during the inflationary scenario may play a fundamental  role in producing the primordial fluctuations \cite{6252602,1126,6252603}. During the warm
inflationary scenario the density perturbations arise from
thermal fluctuations of the inflaton  and dominate over the
quantum ones. In this form,
an essential  condition for warm inflation to occur is the
existence of a radiation component with temperature $T>H$, since the thermal and quantum
fluctuations are proportional to $T$ and $H$,
respectively \cite{warm1,warm2,6252602,1126,6252603,6252604,62526,Moss:2008yb,Ramos:2013nsa}. When the universe heats
up and becomes radiation dominated, inflation ends and the
universe smoothly enters in the radiation
Big-Bang phase. For a comprehensive review of warm
inflation, see Refs. \cite{Berera:2008ar,Ramos:2016coz}.

Alternatively, single-field inflation can de studied using the Hamilton-Jacobi approach \cite{Sayar:2017pam,Sheikhahmadi:2016wyz,Aghamohammadi:2014aca,Kinney:1997ne}.
It is a powerful formulation that allows us to rewrite the equations of motion in an equivalent form assuming that the inflaton itself, and not the cosmic time, is the independent variable. This is possible during any epoch in which the scalar field evolves monotonically with time. Since the Hubble parameter, unlike the inflaton potential, is a geometrical quantity, inflation is described more naturally in a language in which the fundamental
quantity to be considered is $H(\phi)$ rather than $V(\phi)$. For instance,
$H(\phi)\sim exp (\phi)$ corresponds to power-law inflation \cite{Lucchin:1984yf}. Furthermore, this formalism has been adopted
by the Planck collaboration in order to reconstruct the inflaton
potential beyond the slow-roll approximation \cite{Ade:2015lrj}.

Recently, in Ref.\cite{Videla:2016ypa} it was studied a quasi-exponential form for the Hubble rate, given by $H(\phi)=H_{inf} exp\left[ \frac{\phi}{p (\phi+m_p)} \right] $. Under the Hamilton-Jacobi approach, it was obtained an inflaton potential of the form $V(\phi)=\frac{3 m_p^2H^2_{inf}}{8\pi} exp\left[ \frac{2\phi}{p (\phi+m_p)} \right]$. An interesting feature of this potential is that it solves the problem of exit from inflation in comparison to very well known power-law potential. However, the obtained inflaton potential does not present a minimum, which raises the issue of how to address the problem of reheating in this model. However, the author mentioned that this issue may be addressed by the warm inflation scenario. In this way, the main goal of the present work is studied the implications of a concrete warm inflationary model defined by the quasi-exponential potential. In order to describe the dissipative effects, we consider an inflaton decay rate proportional to the Hubble rate, i.e. $\Gamma=3\alpha H$, where $\alpha$ is a constant parameter.

This paper is organized as follows: In the next section, we present
the basic equations of warm inflation. In section \ref{WQE} we study the background and perturbative dynamics of our concrete warm inflationary model. Specifically, we find explicit expressions for the most relevant inflationary observables, such as scalar power spectrum, scalar spectral index and tensor-to-scalar ratio. In addition, we discuss primordial non-Gaussianities of this model, through the computation of the non-linear parameter $f_{NL}$, which will be compared with current bounds imposed by the latest Planck data. Finally, we conclude our work in section \ref{conclu} where we summarize our findings. We work in units where $c=\hbar=1$.

\section{Basics of warm inflation scenario}\label{WI}

\subsection{Background evolution}

We start by considering a spatially flat Friedmann-Robertson-Walker (FRW) universe
containing a self-interacting inflaton scalar field $\phi$ with energy density and pressure
given by $\rho_{\phi}=\dot{\phi}^2/2+V(\phi)$ and $P_{\phi}=\dot{\phi}^2/2-V(\phi)$, respectively,
and a radiation field with energy density $\rho_{\gamma}$. The corresponding Friedmann equations reads
\begin{equation}
H^2=\frac{8\pi}{3m^2_p}(\rho_{\phi}+\rho_{\gamma}),\label{Freq}
\end{equation}
 with $m_p=1.22 \times 10^{19} GeV$ being the Planck mass.

The dynamics of $\rho_{\phi}$ and $\rho_{\gamma}$ is described by the equations \cite{warm1,warm2}
\begin{equation}
\dot{\rho_{\phi}}+3\,H\,(\rho_{\phi}+P_{\phi})=-\Gamma \dot{\phi}^{2},
\label{key_01}%
\end{equation}
and
\begin{equation}
\dot{\rho}_{\gamma}+4H\rho_{\gamma}=\Gamma \dot{\phi}^{2}, \label{key_02}%
\end{equation}
where the  dissipative coefficient $\Gamma>0$ produces the decay of the scalar
field into radiation. Recall that this
decay rate can be assumed  to be a function of the
temperature of the thermal bath $\Gamma(T)$, or a function of the
scalar field $\Gamma(\phi)$, or a function of $\Gamma(T,\phi)$ or
simply a constant.

During warm inflation, the energy density related to
the scalar field predominates  over the energy density of the
radiation field, i.e.,
$\rho_\phi\gg\rho_\gamma$\cite{warm1,warm2,6252602,1126,6252603,6252604,62526,Moss:2008yb}, but even if small when compared to the inflaton energy density
it can be larger than the expansion rate with $\rho_{\gamma}^{1/4}>H$. Assuming thermalization, this translates roughly
into $T>H$, which is the condition for warm inflation to occur.

When $H$, $\phi$, and $\Gamma$ are slowly varying, which is a good
approximation during inflation, the production of radiation becomes quasi-stable, i.e., $\dot{\rho
}_{\gamma}\ll4H\rho_{\gamma}$ and $\dot{\rho}_{\gamma}\ll\Gamma\dot{\phi}^{2}%
$, see Refs.\cite{warm1,warm2,6252602,1126,6252603,6252604,62526,Moss:2008yb}. Then, the equations of motion reduce to
\begin{equation}
3\,H\,(1+Q)\dot{\phi}\simeq -V_{,\phi},
\label{key_01n}%
\end{equation}
where $,\phi$ denotes differentiation with respect to inflaton, and
\begin{equation}
4H\rho_{\gamma}\simeq \Gamma\,\dot{\phi}^{2}, \label{key_02n}%
\end{equation}
where $R$ is the dissipative ratio defined as
\begin{equation}
Q\equiv\frac{\Gamma}{3H}.
\end{equation}

In warm inflation, we can distinguish between two possible scenarios, namely the weak and strong dissipative regimes, defined as $Q\ll 1$ and $Q\gg 1$, respectively. In the weak dissipative regime, the Hubble damping is still the dominant term, however, in the strong dissipative regime, the dissipative coefficient $\Gamma$ controls the damped evolution of the inflaton field.

If we consider thermalization, then the energy density of the radiation field could be written as $\rho_{\gamma}=C_{\gamma}\,T^{4}$, where the constant  $C_{\gamma}=\pi^{2}\,g_{\ast}/30$. Here,   $g_{\ast}$ represents the number
of relativistic degrees of freedom. In the Minimal Supersymmetric Standard Model (MSSM), $g∗ = 228.75$ and $C_{\gamma} \simeq 70$ \cite{62526}. Combining Eqs.(\ref{key_01n}) and (\ref{key_02n}) with $\rho_{\gamma}\propto\,T^{4}$, the temperature of the
thermal bath becomes
\begin{equation}
T=\left[\frac{\Gamma\,V_{,\phi}^2}{36 C_{\gamma}H^3(1+Q)^2}\right]^{1/4}.\label{temp}
\end{equation}

On the other hand, the consistency conditions for the approximations to hold imply that a set of slow-roll conditions must be satisfied for a prolonged period of inflation to take place. For warm inflation, the slow-roll parameters are \cite{26,62526}
\begin{widetext}
\begin{equation}
\epsilon=\frac{m^2_p}{16\pi}\left(\frac{V_{,\phi}}{V}\right)^2,\,\,\,\eta=\frac{m^2_p}{8\pi}\left(\frac{V_{,\phi \phi}}{V}\right),\,\,\,\beta=\frac{m^2_p}{8\pi} \left(\frac{\Gamma_{,\phi}\,V_{,\phi}}{\Gamma\,V}\right),\,\,\,\sigma = \frac{m^2_p}{8\pi} \left(\frac{V_{,\phi}}{\phi V}\right).\label{srparam}
\end{equation}
\end{widetext}

The slow-roll conditions for warm inflation can be expressed as \cite{26,62526,Moss:2008yb}
\begin{equation}
\epsilon \ll 1+Q,\,\,\,\eta \ll 1+Q,\,\,\,\beta \ll 1+Q,\,\,\,\sigma\ll 1+Q\label{srcon}
\end{equation}

When one these conditions is not longer satisfied, either the motion of the inflaton is no
longer overdamped and slow-roll ends, or the radiation becomes comparable to the inflaton energy density. In this way,
inflation ends when one of these parameters become the order of $1+R$.

The number of $e$-folds in the slow-roll approximation, using  (\ref{Freq}) and (\ref{key_01n}), yields
\begin{equation}
N \simeq -\frac{8\pi}{m_p^2}\int_{\phi_{*}}^{\phi_{end}}\frac{V}{V_{,\phi}}(1+Q)d\phi,\label{Nfolds}
\end{equation}
where $\phi_{*}$ and $\phi_{end}$ are the values of the scalar field when the cosmological scales crosses the Hubble-radius and at the end of inflation, respectively.
As it can be seen, the number of $e$-folds is increased due to an extra term
of $(1+Q)$. This implies a more amount of inflation, between these two values of the field, compared to cold
inflation.

\subsection{Cosmological perturbations}

In the warm inflation scenario, a
thermalized radiation component is present with $T>H$, then the inflaton fluctuations
$\delta \phi$ are predominantly thermal instead quantum. In this way, following \cite{1126,62526,Moss:2008yb,Berera:2008ar}, the
amplitude of the power spectrum of the curvature perturbation is given by

\begin{equation}
{\cal{P}_{\cal{R}}}^{1/2}\simeq \left(\frac{H}{2\pi}\right) \left(\frac{3H^2}{V_{,\phi}}\right)\left(1+Q\right)^{5/4}\left(\frac{T}{H}\right)^{1/2},\label{PR}
\end{equation}
where the normalization has been chosen in order to recover the standard cold inflation result when $Q\rightarrow 0$ and $T \simeq H$.

By the other hand, the scalar spectral index $n_s$ to leading order in the slow-roll approximation, is given by \cite{62526,Moss:2008yb}
\begin{equation}
n_s=1+\frac{d\ln{\cal{P}_{\cal{R}}}}{d\ln k}\simeq 1-\frac{(17+9Q)}{4(1+Q)^2}\epsilon-\frac{(1+9Q)}{4(1+R)^2}\beta+\frac{3}{2(1+Q)}\eta.\label{nsw}
\end{equation}

Regarding to tensor perturbations, these do not couple to the thermal background, so gravitational waves are only generated by quantum fluctuations, as
in standard inflation \cite{Ramos:2013nsa}. However, the tensor-to-scalar ratio $r$ is modified with respect to standard cold inflation, yielding \cite{Berera:2008ar}
\begin{equation}
r\simeq \left(\frac{H}{T}\right)\frac{16\epsilon}{(1+Q)^{5/2}}.\label{rwi}
\end{equation}
We can see that warm inflation predicts a tensor-to-scalar ratio suppressed by a factor $(T/H)(1 + Q)^{
5/2} > 1$ compared with standard cold inflation.

When a specific form of the scalar potential and the dissipative coefficient $\Gamma$ are considered, it is possible to study the
background evolution under the slow-roll regime and the primordial perturbations in order to
test the viability of warm inflation.

\subsection{Non-Gaussianities in warm inflation}

Due to the existence of a wide range of inflationary universe models it is important to
discriminate between them. Non-Gaussianities is one of the features that can help us in this direction.
In fact, non-Gaussian statistics (such as the bispectrum) provides us with a
powerful tool to discriminate between different mechanisms for generating
the curvature perturbation \cite{Bartolo:2004if}. But this feature not only well help us to discriminate between
inflationary scenarios, but also, measurements (including an upper bound) of non-Gaussianities
of primordial fluctuations are expected to have the potential to rule out many of inflationary
models that have been put forward \cite{Chen:2010xka,Chen:2006nt}.

It has been notice that a single field, slow-roll inflationary scenarios are known to produce
negligible non-Gaussianities \cite{Maldacena:2002vr}, there exist now a
variety of models available in the literature which may predict an observable signature. One
important referent of this situation is warm inflation. The reason of this is due that warm
inflation could be seen as a model which is analogous to a multi-field inflation scenario, which is well know that can produce large non-Gaussianities which can be observed \cite{Bastero-Gil:2014raa}. The constraint on primordial non-Gaussianities, which is parameterized by the non-linear parameter $f_{NL}$, is currently obtained from CMB measurements \citep{Ade:2015ava}.

In Ref.\cite{Zhang:2015zta}, the authors obtained and analytical expression for the non-linear parameter $f_{NL}$ in the warm inflation scenario, by using the $\delta N$ formalism under slow-roll approximation. For an inflaton decay rate having an inflaton field dependence, i.e, $\Gamma=\Gamma(\phi)$, the expression obtained for $f_{NL}$ was given by
\begin{equation}
-\frac{3}{5}f_{NL}=-\frac{\epsilon}{1+Q}+\frac{\eta}{2(1+Q)}+\frac{Q \epsilon}{(1+Q)^2}-\frac{Q \beta}{2(1+Q)^2},\label{fnl}
\end{equation}
where $\epsilon$, $\beta$, and $\eta$ are the slow-roll parameters already defined in Eq.(\ref{srparam}). For the two concrete examples
the authors studied, quartic chaotic and the
hilltop models, they found that the non-linear parameters for both cases are consistent with current bounds imposed by Planck.\\

In the following we will study how an inflaton decay rate proportional to Hubble rate, i.e. $\Gamma=3\alpha H$, with $\alpha$ being a dimensionless parameter, influences the inflationary dynamics for the quasi-exponential potential. We will study the dynamics under slow-roll approximation without assuming any dissipative regime of warm inflation in particular. In addition we also study the predictions of our model regarding primordial non-Gaussianity, trough the non-linear parameter $f_{NL}$.

\section{Dynamics of warm quasi-exponential inflation}\label{WQE}

Here we analyse in detail a concrete warm inflationary model defined by the following inflaton potential, Hubble rate and inflaton decay rate
\begin{eqnarray}
\label{V2}
V(\phi) & = & \frac{3 m_p^2 H_{inf}^2}{8 \pi} exp\left[ \frac{2 \phi}{p (\phi+m_p)} \right] \\
\label{H}
H(\phi) & = & H_{inf} exp\left[ \frac{\phi}{p (\phi+m_p)} \right] \\
\label{Gamma2}
\Gamma(\phi) & = & 3 \alpha H(\phi)
\end{eqnarray}
respectively. The model is characterized by three free parameters $p, \alpha, h=H_{inf}/m_p$, and we
study the model for a generic parameter $\alpha$ without making a distinction between weak and strong dissipative regime.
Combining the cosmological equations the temperature $T$ as a function of the scalar field $\phi$ is found to be
\begin{equation}
T(\phi)=\left( \frac{\alpha m_p^2 V_{,\phi}^2}{1120 \pi (1+\alpha)^2 V} \right)^{1/4}
\end{equation}
In the following we introduce for convenience the dimensionless parameter $y=\phi/m_p$.
The end of inflation $y_{end}$ is determined by the condition $\eta_{end}=1$, where the slow-roll parameter $\eta$
is computed to be
\begin{equation}
\eta(y) = \frac{(1-p)-p y}{2 \pi p^2 (1+\alpha) (1+y)^4},
\end{equation}
while observables are evaluated at $y_*$, computed using the number of $e$-folds
\begin{equation}
N = \frac{8 \pi (1+\alpha)}{m_p^2} \int_{\phi_{end}}^{\phi_*} d\phi \frac{V}{V_{,\phi}}.
\end{equation}
Using the general formulas of the previous section, for the model at hand the power spectrum, the scalar spectral index and the tensor-to-scalar
ratio are computed to be
\begin{widetext}
\begin{eqnarray}
\mathcal{P}_{\mathcal{R}} & = & \sqrt{\frac{2}{\pi}} \left( \frac{3}{35} \right)^{1/4} p^{3/2} \alpha^{1/4} (1+\alpha)^2 (1+y_*)^3 h^{3/2} exp\left[ \frac{3 y_*}{2 p (1+y_*)} \right] \\
n_s & = &  1 - \frac{3+6 p (1+y_*)}{8 \pi p^2 (1+\alpha)^2 (1+y_*)^4} \\
r & = & 8 \sqrt{\frac{2 h}{\pi}} \left( \frac{35}{3} \right)^{1/4} p^{-3/2} \alpha^{-1/4} (1+\alpha)^{-3} (1+y_*)^{-3} exp\left[ \frac{y_*}{2 p (1+y_*)} \right]
\end{eqnarray}
\end{widetext}
respectively.

\begin{figure}[ht!]
\centering
\hspace{-1 in}
\includegraphics[scale=0.40]{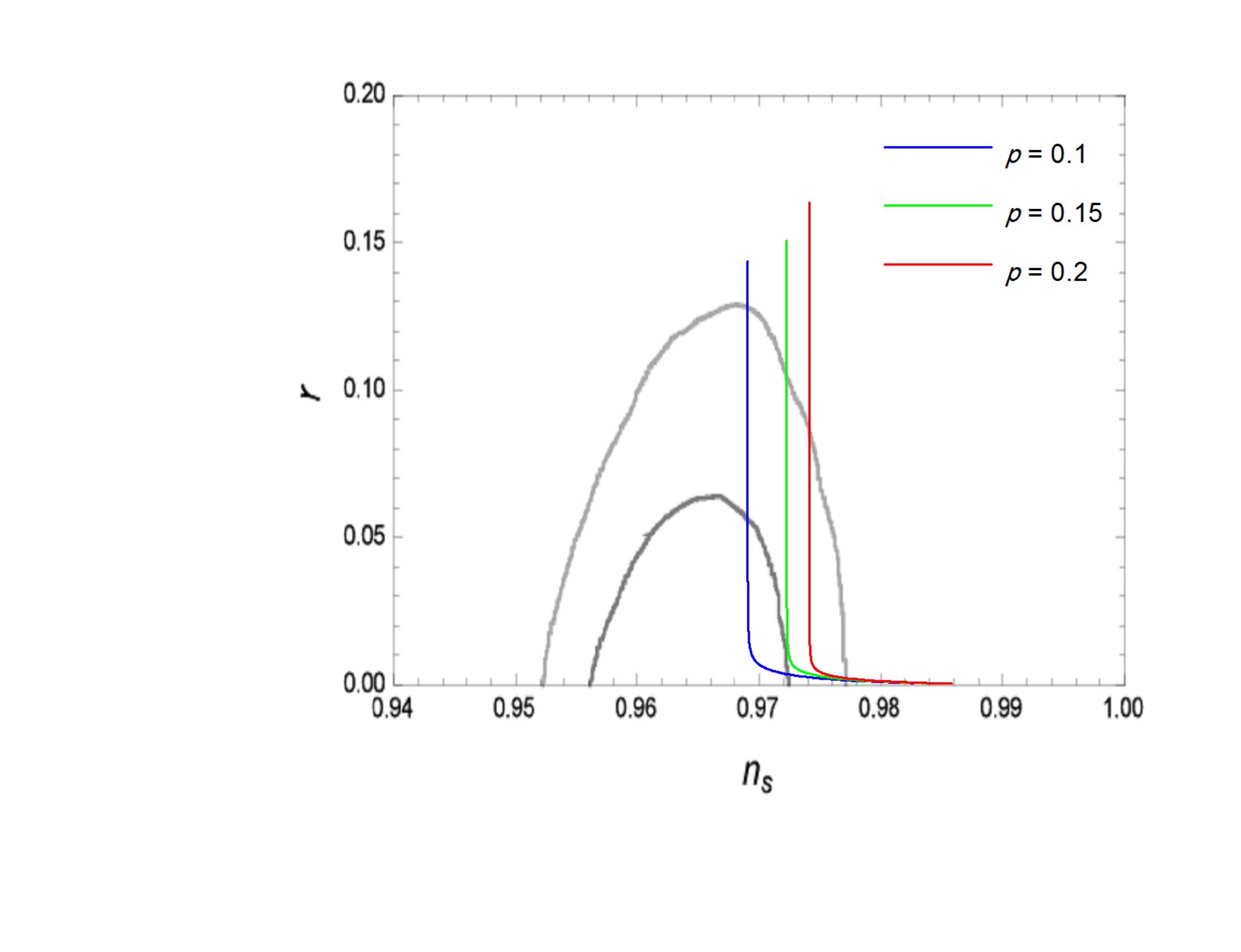}
{\vspace{-0.5 in}}
\caption{Allowed contours at the 68 and 95 $\%$ C.L., from the latest Planck data \cite{Ade:2015lrj} and theoretical predictions in the plane $r$ versus $n_s$ for $N=60$ and for three cases: $p=0.1$ (blue line), $p=0.15$ (green line), and $p=0.2$ (red line).}
\label{fig:1} 	
\end{figure}

First we use the COBE normalization $\mathcal{P}_{\mathcal{R}}=2 \times 10^{-9}$ as a constraint to express $h$ in terms of $p,\alpha$. Then $r$ and $n_s$ for a given number of e-folds
are certain functions of $p$ and $\alpha$. We fix $N=60$ and consider three cases $p=0.1, 0.15, 0.2$. For each case we plot $r$ versus $n_s$ in the same plot with the allowed contour plots, as is shown in Fig.\ref{fig:1}. The theoretical prediction lies inside the allowed region when $\alpha$ takes values in the following range:\\
For $p=0.1$
\begin{equation}
6.99 \times 10^{-6} < \alpha < 4.45 \times 10^{-1}
\end{equation}
and accordingly
\begin{equation}
2.12 \times 10^{-9} > h > 3.34 \times 10^{-10}
\end{equation}
For $p=0.15$
\begin{equation}
5.66 \times 10^{-6} < \alpha < 2.61 \times 10^{-1}
\end{equation}
and accordingly
\begin{equation}
3.48 \times 10^{-8} > h > 5.58 \times 10^{-9}
\end{equation}
For $p=0.2$
\begin{equation}
6.0 \times 10^{-6} < \alpha < 1.55 \times 10^{-1}
\end{equation}
and accordingly
\begin{equation}
1.30 \times 10^{-7} > h > 2.30 \times 10^{-8}
\end{equation}

\begin{figure}[ht!]
\centering
\hspace{0 in}
\includegraphics[scale=0.40]{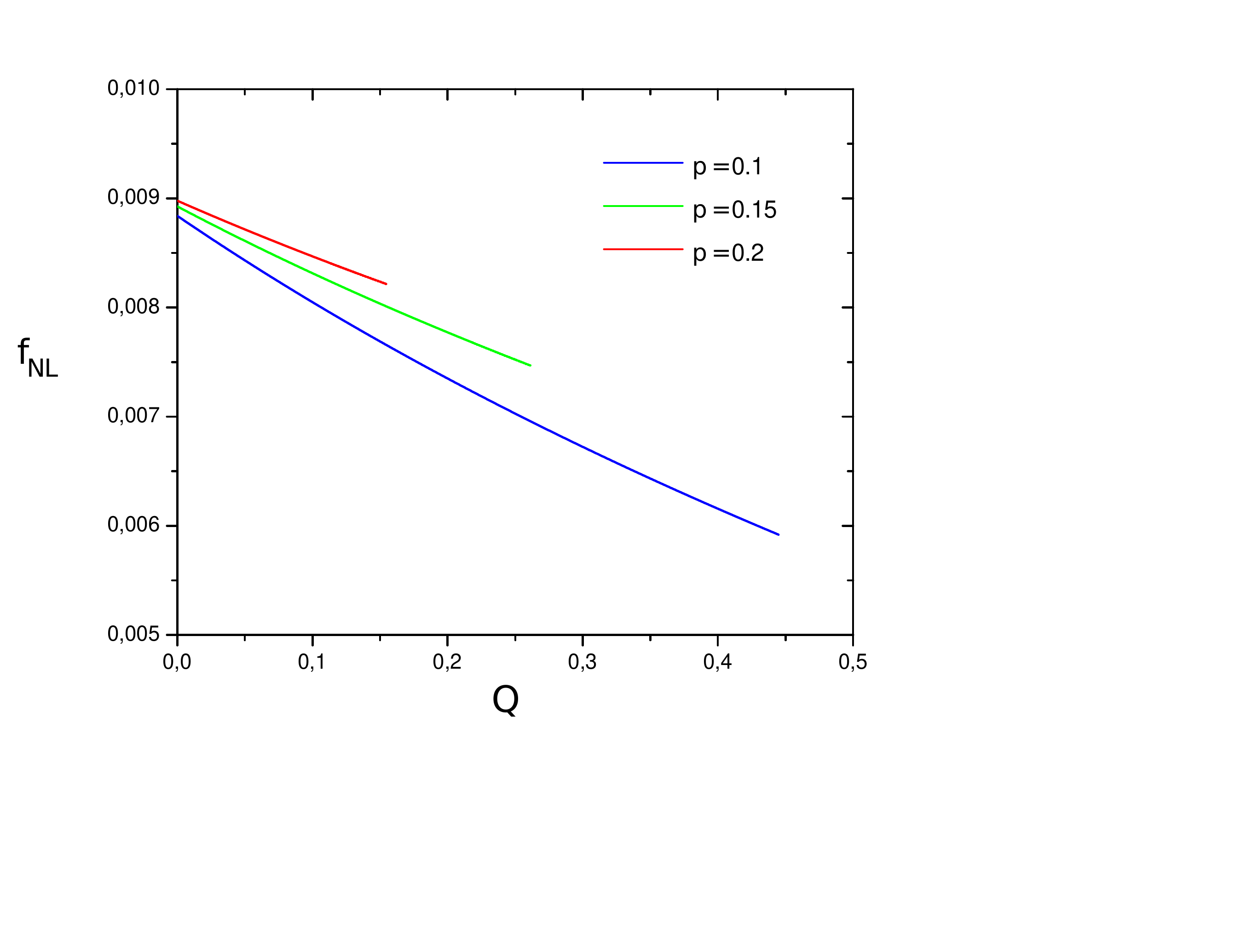}
\vspace{-0.5 in}
\caption{Non-linear parameter $f_{NL}$ as a function of $Q=\alpha$ for $N=60$ and for three cases: $p=0.1$ (blue line), $p=0.15$ (green line), and $p=0.2$ (red line).}
\label{fig:2} 	
\end{figure}

It is interesting to mention that the theoretical predictions  $h=H_{inf}/m_p$ at the time when the cosmological scales exit the Hubble radius during inflation are consistent with the lower bound for the Hubble rate at the same time set by Planck, which is given by $H_{*}/m_p< 3\textup{.}65\times 10^{-5}$ at $95\%$ C.L.

In addition, as a consistency test, we have checked that in all cases for the allowed ranges of the free parameters of the model, the condition for warm inflation $T > H$ is satisfied.

Finally, regarding primordial non-Gaussianities, and using the general formulas of the previous section, the prediction for the non-linear parameter $f_{NL}$ is found to be
\begin{equation}
f_{NL}=\frac{10\,p\,(1+\alpha)(1+y_{*})-5\alpha}{24\,p^2\,\pi(1+\alpha)^2(1+y_{*})^4},\label{fnlmodel}
\end{equation}
which is evaluated at $y_{*}$, through de number of $e$-folds. By fixing the number of $e$-folds to $N=60$, we plot the non-linear parameter $f_{NL}$ as a function of the dissipation strength of warm inflation $Q=\alpha$ for three cases $p=0.1$, $p=0.15$, and $p=0.2$, as it is depicted in Fig.\ref{fig:2}. For each value of $p$, we consider the allowed range for $\beta$ already obtained. From Fig.\ref{fig:2} we observe that the magnitude of $f_{NL}$ decreases as $Q=\alpha$ increases. In this way, for each value of $p$, the effects of non-Gaussianities are not significant, being $\mathcal{O}(10^{-2})$, when the cosmological scales cross the Hubble-radius at $N=60$. The current observational value for $f_{NL}$ in warm inflation imposed by the latest Planck observations, lies in the range $f_{NL}^{\textup{warm}}=-23\,\pm\,36$ at $68\%$ C.L. The predictions of our model, consisting in canonical single field in warm inflation, yields a small but positive value for $f_{NL}$, being marginally consistent with the negative central value from the Planck collaboration. However, as it has been suggested in Ref.\cite{Zhang:2017syn}, warm inflation driven by a non-canonical field may generate a larger amount of non-Gaussianities than the canonical case. We hope to be able to address that issue in a future work.

\section{Conclusions}\label{conclu}

In the standard cold inflation scenario, a quasi-exponential form
for the Hubble rate studied under the Hamilton-Jacobi approach, yields an inflaton potential of the form $V(\phi) = \frac{3 m_p^2H^2_{inf}}{8\pi} exp\left[ \frac{2\phi}{p (\phi+m_p)} \right] $, which solves the problem of exit from inflation in comparison to very well known power-law potential. However, the obtained inflaton potential does not present a minimum, which raises the issue of how to address the problem of reheating in
this model. In order to address this problem, in the present work we have studied the implications of a concrete warm inflationary model defined by the quasi-exponential potential, and an inflaton decay rate proportional to the Hubble rate, $\Gamma=3\alpha H$. In total, the model is characterized by three free parameters, namely $h=H_{inf}/m_p$ and $p$ from the potential, and $\alpha$ from the inflaton decay rate. Contact between the predictions of the model and observations is made by computing the power spectrum, the scalar spectral index as well as the tensor-to-scalar ratio. The COBE normalization is first used as a constraint to express the inflationary scale in terms of the other two parameters of the model. Then on the $r-n_s$ plane we show both the theoretical curves and the allowed Planck's contour plots. Requiring that the theoretical curves lie within the observationally allowed region we were able to constrain the parameters of the model. Finally, primordial non-Gaussianities and the non-linear parameter $f_{NL}$ are also briefly discussed. We found that the effects of non-Gaussianities are not significant and also that the value predicted for $f_{NL}$ lies within the range imposed by the latest Planck data.  As we mentioned, warm inflation driven by a non-canonical field may generates a large amount of non-Gaussianities in comparison to canonical case. In this direction, we left the consequences of studying a warm quasi-exponential inflation with a non-canonical field as a future work.


\begin{acknowledgments}
N.V. was supported by Comisi\'on Nacional
de Ciencias y Tecnolog\'ia of Chile through FONDECYT Grant N$^{\textup{o}}$
3150490. Additionally, N.V. would like to express his gratitude to the
Instituto Superior T\'ecnico of Universidade de Lisboa
for its kind hospitality during the initial stage of
this work. G. P. thanks the Funda\c c\~ao para a Ci\^encia e Tecnologia (FCT), Portugal, for the financial support to the Multidisciplinary Center for Astrophysics (CENTRA),  Instituto Superior T\'ecnico,  Universidade de Lisboa,  through the Grant No. UID/FIS/00099/2013.
\end{acknowledgments}


\end{document}